\documentclass[a4paper,aps,floatfix,pra,showpacs,twocolumn]{revtex4}

\usepackage{amsfonts,amsmath,amssymb}
\usepackage{graphicx}
\usepackage[utf8x]{inputenc}
\usepackage{microtype}
\usepackage{textcomp}
\usepackage{units}
\usepackage{wrapfig}
\usepackage{hyperref}

\newcommand{\degree}{\ensuremath{^\circ}}%
\newcommand{\ie}{i.\,e.}%
\newcommand{\cosTwoTheta}[1]{\ensuremath{\left<\cos^2\theta\right>=#1}}
\newcommand{\classTheta}[1]{\ensuremath{\theta_\text{cl}=#1\,\degree}}
\newcommand{\keV}[1]{\ensuremath{\text{#1~keV}}}
\newcommand{\micrometer}[1]{\ensuremath{\text{#1~\textmu{m}}}}

\newcommand{\picometer}[1]{\ensuremath{\text{#1~pm}}}

\newlength{\figwidth}
\newlength{\figwidthsmall}
\newcommand{\mueff}{\ensuremath{\mu_\text{eff}}}

\begin{document}

\title[State- and conformer-selected beams of aligned and oriented molecules for ultrafast
diffraction studies]{State- and conformer-selected beams of aligned and oriented molecules \\ for
   ultrafast diffraction studies}

\author{Frank Filsinger$^1$}%
\author{Gerard Meijer$^1$}%
\author{Henrik Stapelfeldt$^2$}%
\author{Henry N.\ Chapman$^3$}%
\author{Jochen K\"upper$^1$}%
\email{jochen@fhi-berlin.mpg.de}%
\affiliation{${}^1$~Fritz-Haber-Institut der Max-Planck-Gesellschaft, Faradayweg 4-6, 14195 Berlin,
   Germany \\
   ${}^2$~University of Aarhus, Department of Chemistry and Interdisciplinary Nanoscience
   Center (iNANO), 8000 Aarhus C, Denmark \\
   ${}^3$~Center for Free Electron Laser Science, DESY and University of Hamburg, 22607 Hamburg,
   Germany }%


\date{\today}

\begin{abstract}\noindent%
   The manipulation of the motion of neutral molecules with electric or magnetic fields has seen
   tremendous progress over the last decade. Recently, these techniques have been extended to the
   manipulation of large and complex molecules. In this article we introduce experimental approaches
   to the manipulation of large molecules, \ie, the deflection, focusing and deceleration using
   electric fields. We detail how these methods can be exploited to spatially separate quantum
   states and how to select individual conformers of complex molecules. We briefly describe
   mixed-field orientation experiments made possible by the quantum-state selection. Moreover, we
   provide an outlook on ultrafast diffraction experiments using these highly controlled samples.
\end{abstract}
\maketitle

\section{Introduction}
\label{sec:intro}

\subsection{Manipulation of molecular beams with electric and magnetic fields}
\label{sec:intro:fields}

\subsubsection{Deflection of polar molecules}

By expanding atoms or molecules from a reservoir at high pressure into vacuum a so called atomic or
molecular beam is created. In such beams the molecules' intrinsic properties can be investigated
under collision-free conditions, independent from interactions with other species. A century ago,
when such beams were initially investigated~\cite{Dunoyer:ComptRend152:592}, laser-based
quantum-state-selective detection techniques were still lacking. In 1921 Stern proposed that the
trajectories of silver atoms on their way to the detector could be characteristically altered,
depending on their quantum state, when the atomic beam was exposed to an inhomogeneous magnetic
field~\cite{Stern:ZP7:249}. In a ground-breaking experiment, Gerlach and Stern demonstrated in
1922~\cite{Gerlach:ZP9:349} that indeed quantum-state selectivity could be achieved in the detection
process by sorting different quantum states \emph{via} space quantization, a concept that has been
extensively used ever since. The possibility to deflect polar molecules in a molecular beam with
electric fields was conceived at the same time. It was first theoretically described by Kallmann and
Reiche in 1921~\cite{Kallmann:ZP6:352}\footnote{In fact, Stern states in a footnote of his initial
   paper on space quantization~\cite{Stern:ZP7:249} that its publication was motivated by Kallmann
   and Reiche's article, of which he had received the galley proofs.} and later experimentally
demonstrated by Wrede -- a graduate student of Stern -- in 1927~\cite{Wrede:ZP44:261}.

As early as 1926, Stern suggested that the technique could be used for the quantum-state separation
of small diatomic molecules at low temperatures~\cite{Stern:ZP39:751}. Over the years, various
experimental geometries were designed to create strong field gradients on the beam axis in order to
efficiently deflect particles. In 1938/1939 Rabi introduced the molecular beam magnetic resonance
method, by using two deflection elements of oppositely directed gradients in succession, to study
the quantum structure of atoms and molecules~\cite{Rabi:PR53:318,Rabi:PR55:526}. In his setup, the
deflection of particles caused by the first magnet was compensated by a second magnet such that the
particles reached the detector on a sigmoidal path. If in between the two magnets a transition to a
different quantum state was induced, this compensation was incomplete and a reduction of the
detected signal could be observed. For more details on these historic experiments we refer to
Ramsey's classical textbook on ``Molecular Beams''~\cite{Ramsey:MolBeam:1956}. Since these early days of molecular beam deflection experiments,
the deflection technique has been widely used as a tool to determine dipole moments and
polarizabilities of molecular systems ranging from diatomics~\cite{Wrede:ZP44:261} to
clusters~\cite{Moro:Science300:1265, deHeer:arXiv0901:4810} to large
biomolecules~\cite{Broyer:PhysScr76:C135}.

\subsubsection{Focusing and deceleration of molecules in low-field-seeking quantum states}

Whereas deflection experiments allow for the spatial dispersion of quantum states, they do not
provide any focusing of the molecular beam. For small molecules in eigenstates whose energy
increases with increasing field strength, so-called low-field-seeking (lfs) states, focusing was
achieved using multipole focusers. Both magnetostatic~\cite{Friedburg:Nat38:159,
   Bennewitz:ZP139:489} and electrostatic~\cite{Bennewitz:ZP141:6} devices were developed in the
early 1950s by Paul's group in Bonn. Independently, an electrostatic quadrupole focuser, \ie, a
symmetric arrangement of four cylindrical electrodes around the beam axis that are alternately at
positive and negative voltage, was built in 1954/55 by Gordon, Zeiger and Townes in New York to
create the population inversion of ammonia molecules for the first demonstration of the
MASER~\cite{Gordon:PR95:282, Gordon:PR99:1264}. Using several multipole focusers in succession and
interaction regions with electromagnetic radiation in between them, many setups were developed to
unravel the quantum structure of atoms and molecules -- very similar to Rabi's molecular beam
magnetic resonance method. About ten years after the invention of the multipole focusing technique,
molecular samples in a single rotational state were used for state-specific inelastic scattering
experiments by the Bonn group~\cite{Bennewitz:ZP177:84} and, shortly thereafter, for reactive
scattering studies~\cite{Brooks:JCP45:3449, Beuhler:JACS88:5331}. In the following decades,
multipole focusers were extensively used to study steric effects in gas-phase reactive scattering
experiments~\cite{Parker:ARPC40:561, Stolte:StateSelectedScattering:1988:1and2}. The preparation of
oriented samples of state-selected molecules using electrostatic focusers was also essential for the
investigation of steric effects in gas-surface scattering~\cite{Kuipers:Nature334:420} and
photodissociation~\cite{Rakitzis:Science303:1852} experiments. Variants of multipole focusing setups
were implemented in many laboratories all over the world and yielded important information on stable
molecules, radicals, and molecular complexes.

Finally, in 1999 the so-called Stark decelerator was realized~\cite{Bethlem:PRL83:1558}, allowing
the same control over the forward velocities of molecules in lfs states. This technique was used to
confine small molecules in storage rings~\cite{Crompvoets:Nat411:174} and
static~\cite{Bethlem:Nature406:491} and dynamic traps~\cite{Veldhoven:PRL94:083001}. Recently, the
``decelerator on a chip'' -- a miniaturized version of the Stark decelerator -- has been
implemented~\cite{Meek:Science324:1699}. Detailed accounts of the field of Stark deceleration have
been given elsewhere~\cite{Meerakker:NatPhys4:595, Bell:MP107:99,
   Meerakker:SlowingAndTrapping:2009}.


\subsubsection{Focusing and deceleration of molecules in high-field- seeking quantum states}

Large or heavy molecules have small rotational constants and, as a consequence, a high density of
rotational states. Coupling between closely spaced states of the same symmetry turns lfs states into
hfs states already at relatively weak electric field strengths (compared to the field strengths that
are required for efficient focusing). In order to focus molecules in these states, a maximum of the
electric field in free space would have to be created. Since Maxwell's equations do not allow for
the creation of a 3D maximum with static fields alone~\cite{Ketterle:APB54:403,
   Wing:ProgQuantElectr8:181}, static multipole fields cannot be applied to focus molecules in hfs
states. The situation is analogous to charged particle physics: charged particles also cannot be
confined with static potentials alone. This focusing problem for ions was solved when Courant,
Livingstone, and Snyder introduced the principle of ``alternating gradient (AG) focusing'' in the
1950s~\cite{Courant:AnnPhys3:1, Courant:PhysRev88:1190}. The basic idea is to create an array of
electrostatic lenses that focus the particles along one transverse coordinate while defocusing them
along the perpendicular transverse axis. Alternating the orientation of these fields at the
appropriate frequency results in a net focusing force along both transverse coordinates. This
principle is exploited to confine ions, for instance, in quadrupole mass
filters~\cite{Paul:RMP62:531, Paul:ZNATA8:448}, in Paul traps~\cite{Paul:FWV:1958, Paul:RMP62:531},
and in virtually all particle accelerators. The application of AG focusing to neutral polar
molecules was first proposed by Auerbach, Bromberg, and Wharton~\cite{Auerbach:JCP45:2160} and
experimentally demonstrated by Kakati and Lain\'e for ammonia molecules in hfs
states~\cite{Kakati:PLA24:676, Kakati:PLA28:786, Kakati:JPE4:269}. Later, the diatomic
KF~\cite{Guenther:ZPCNF80:155, Luebbert:CPL35:210} and ICl~\cite{Luebbert:JCP69:5174} molecules were
also focused. More recently, slow ammonia molecules were guided from an effusive source using a bent
AG focuser~\cite{Junglen:PRL92:223001}, even though molecules in lfs and hfs states could not be
distinguished because the detection process was not state selective. Furthermore, CaF molecules have
been guided using a 1~m-long straight AG focuser~\cite{Wall:PRA80:043407}. Besides the AG focusing
technique, various alternative approaches were implemented to focus molecules in hfs states, such as
exploiting the fringe fields of ring-like electrode structures~\cite{AlAmiedy:PL66A:94}, the fields
created by crossed wires~\cite{Laine:PLA34:144}, or the fields created by coaxial
electrodes~\cite{Helmer:JAP31:458, Chien:CP7:161,Loesch:CP207:427, Loesch:PRL85:2709}. Most of these
methods, however, were only used for proof-of-principle experiments and did not find further
applications.

The first attempt to manipulate the forward velocity of molecules in hfs states was reported in the
1960s, when the group of Wharton at the University of Chicago set up an 11~m-long machine to
accelerate LiF molecules~\cite{Wolfgang:SciAm219:44, Bromberg:thesis:1972}. While these early
experiments were unsuccessful and stopped after the PhD student had finished his thesis, a
decelerator design that exploits the AG principle for transverse confinement of the molecules was
successfully implemented in 2002~\cite{Bethlem:PRL88:133003}, inspired by the successful
deceleration of small molecules with the Stark decelerator. So-called AG decelerators were used to
decelerate CO~\cite{Bethlem:PRL88:133003, Bethlem:JPB39:R263}, YbF~\cite{Tarbutt:PRL92:173002}, and
benzonitrile~\cite{Wohlfart:PRA77:031404} molecules in hfs quantum states and OH radicals in both
hfs and lfs states ~\cite{Wohlfart:PRA78:033421, Wohlfart:thesis:2008}. In these first experiments
on high-field-seeking molecules, up to 30\,\% of the kinetic energy was removed, but so far it has
not been possible to decelerate molecules to velocities that are small enough for trapping in
stationary traps.

AC trapping of para-ND$_3$ in the hfs component of its ground rotational state (J$_\text{K}=1_1$)
was achieved by decelerating the molecule in a lfs state with a conventional Stark decelerator and
subsequently transferring the population to the hfs state using microwave
radiation~\cite{Veldhoven:PRL94:083001}.

Moreover, high-frequency AC fields have also been used for the deflection, focusing, and
deceleration of neutral molecules, and these methods are generally applicable to molecules in all,
dc lfs and hfs, states. Strong laser fields have been used to deflect and
focus~\cite{Stapelfeldt:PRL79:2787,Zhao:PRL85:2705} and to decelerate~\cite{Fulton:PRL93:243004} a
fraction of the molecules in a beam. Alternatively, the focusing of molecules with microwave fields
has been demonstrated recently~\cite{Odashima:PRL104:253001}.

\subsection{Large neutral molecules in the gas phase}
\label{sec:intro:large_molecules}

During the last decades, the properties of biomolecules in the gas phase have been studied in ever
greater detail~\cite{EPJD20:Biomolecules, PCCP6:Biomolecules, Vries:ARPC58:585}. Although the study
of biomolecules outside of their natural environment was met with skepticism in the beginning,
spectroscopic studies on isolated species in a molecular beam have proven to be very powerful for
understanding the molecules' intrinsic properties and for benchmarking theoretical calculations.
Moreover, the molecule's native environment can be partly mimicked by adding solvent molecules one
by one~\cite{Vries:ARPC58:585, Kim:JPC100:7933, Piuzzi:CP270:205, Blom:JPCA111:7309}.

Even in the cold environment of a molecular beam, biomolecules exist in various conformational
structures~\cite{Suenram:JACS102:7180, Rizzo:JCP83:4819}. In many cases, the individual conformers
are identified via their different electronic spectra~\cite{Rizzo:JCP83:4819, Nir:Nature408:949}.
Structural information on the individual conformers can be deduced from, for instance,
multiple-resonance techniques, which yield conformer-specific infrared
spectra~\cite{Snoek:CPL321:49, Bakker:PRL91:203003}. Moreover, one can exploit the different angles
between vibrational transition moments and the permanent dipole moments of oriented
molecules~\cite{Dong:Science298:1227}. Quadrupole coupling constants, determined by means of
Fourier-transform microwave spectroscopy~\cite{Lesarri:ACIE43:605}, or permanent dipole moments,
deduced from the rotationally resolved spectra~\cite{Reese:JACS126:11387, Filsinger:PCCP10:666} are
also conformer specific.

The preparation of conformer-selected samples of biomolecules could enable a new class of
experiments to be performed on these systems. For charged species, the separation of structurally
different molecules has been demonstrated using ion mobility in drift
tubes~\cite{Helden:Science267:1483, Jarrold:PCCP9:1659}. For neutral molecules it has been
demonstrated that the abundance of the conformers in the beam can be partly influenced by selective
over-the-barrier excitation in the early stage of the expansion~\cite{Dian:Science296:2369} or by
changing the carrier gas~\cite{Erlekam:PCCP9:3783}. Both methods for neutrals, however, are neither
generally applicable nor able to specifically select conformers.

\begin{figure}
   \centering
   \includegraphics[width=\figwidth]{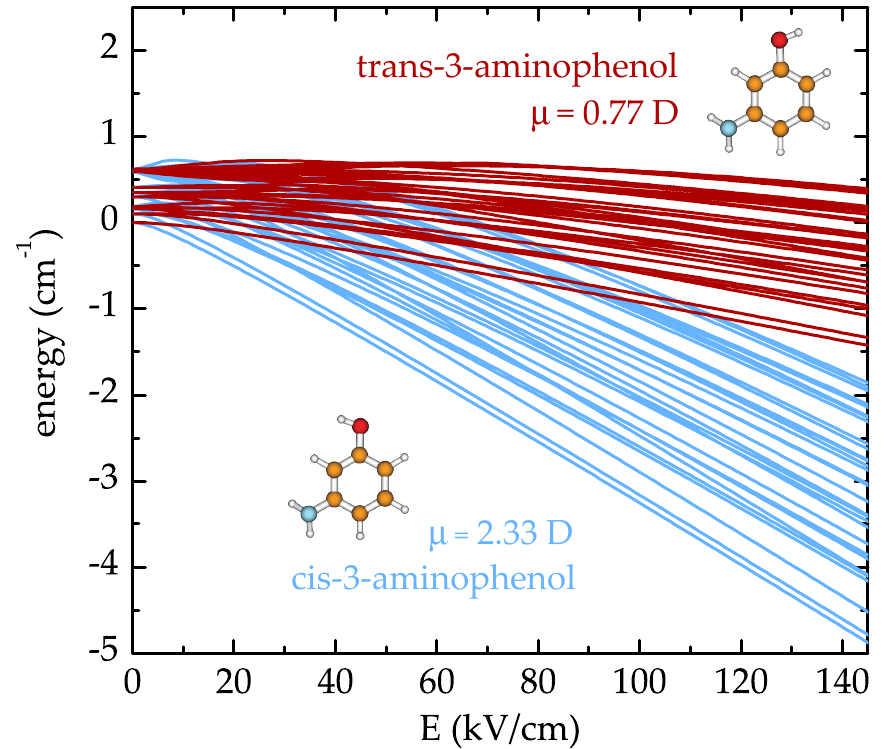}
   \caption{Molecular structures, dipole moments, and energies of the lowest rotational states of
      cis- and trans-3-aminophenol as a function of the electric field strength (reproduced from
      reference~\cite{Filsinger:ACIE48:6900}).}
   \label{fig:3AP:starkshift}
\end{figure}

Spatial separation of conformers can be achieved by exploiting their specific interaction with
electric fields. All conformers of a molecule have the same mass and the same connectivities between
the atoms (primary structure), but often differ by their dipole moments, which are largely
determined by the orientations of the functional groups in the molecular frame, \ie, by the folding
pattern (secondary structure). These different dipole moments lead to different Stark shifts of the
rotational energy levels in an electric field, as shown in \autoref{fig:3AP:starkshift} for the
prototypical large molecule 3-aminophenol (3AP). The force that a molecule experiences in an
electric field is determined by its effective dipole moment $\mu_{\text{eff}}$, which is given by
the negative slope of the Stark curve. From \autoref{fig:3AP:starkshift} it is obvious that the two
conformers of 3-aminophenol will experience different forces in an electric field, which can be
exploited to spatially separate them (\emph{vide infra}).

Such conformer-selected samples are expected to benefit a variety of future applications such as
tomographic imaging experiments~\cite{Itatani:Nature432:867} or ultrafast dynamics studies on the
ground-state potential energy surface. For ultrafast electron and X-ray diffraction
experiments~\cite{Williamson:Nature386:159, Siwick:Science302:1382, Chapman:NatPhys2:839} aiming at
the ``molecular movie'', \ie, measuring chemical processes with spatial and temporal atomic
resolution (a few picometers and femtoseconds, respectively), the preparation of conformer-selected
samples might be crucial, as we will see in the remainder of this article.

\subsection{X-ray diffractive imaging of molecular ensembles}

Here we discuss the possibility of ultrafast diffraction experiments of controlled samples in a
molecular beam. Clearly, X-ray crystallography is at the very heart of structural biology. However,
many biological molecules do not crystallize and many cannot easily be purified. The new X-ray Free
Electron Laser (XFEL) light sources~\cite{Altarelli:XFEL-TDR:2006, Emma:NatPhoton} promise the
possibility to obtain single-molecule diffraction images of large molecules in the gas phase. This
could, for example, help biology to obtain structural information on the large number of
un-crystallizable proteins and other difficult systems~\cite{Huldt:JStructBio144:219}.

The ability to determine the structure of individual biological molecules -- using XFEL radiation --
without the need for purification and crystallization would, therefore, constitute a fundamental
breakthrough for structural biology. However, the proposed experiments for large molecules rely on
the recording of a detectable diffraction pattern from a single molecule in order to be able to
classify and average images from multiple shots~\cite{Spence:PRL92:198102}. It is not \emph{a
   priori} clear whether it is possible to obtain such a single-molecule diffraction image at all,
especially at atomic resolution. Calculations show that the X-ray pulse must be short enough to only
probe the molecule at times before it is converted into a plasma~\cite{Neutze:Nature406:752}.
Moreover, the scattering signal must be large enough to be clearly detectable above all sources of
noise, including scattered light, electric noise, and background signal.

In order to test the feasibility of single-particle diffraction of such large systems, we propose a
bottom-up approach: one would perform large angle diffraction imaging on molecules containing tens
of atoms in order to explore the technical challenges of such experiments. Recently, it was
theoretically investigated which structural information can, in principle, be extracted from the
X-ray diffraction patterns of aligned samples of small symmetric top
molecules~\cite{Ho:JCP131:131101, Pabst:PRA81:043425}. It is clear that for such small species the
signals from many molecules must be averaged. Therefore, it is important to provide samples which
are dense and as clean and defined as possible to allow for experimental averaging over multiple
X-ray pulses and successive image averaging. Here, the existence of the above-mentined isomers turns
out to be a real problem. All isomers will yield individual diffraction images that cannot be
averaged over as this would obscure the structural information. Instead, the structural isomers must
be spatially separated and only a single conformer must enter the imaging system. This isomer
separation can be achieved by the different experimental approaches described in this article.
Moreover, this selection will intrinsically provide samples with most population in the lowest
rotational states, which can be aligned and oriented especially well using strong laser or dc
electric fields (\emph{vide infra}).

\section{Experimental results} 
\label{sec:experimental-results}

\begin{figure}
   \centering%
   \includegraphics[width=\figwidth]{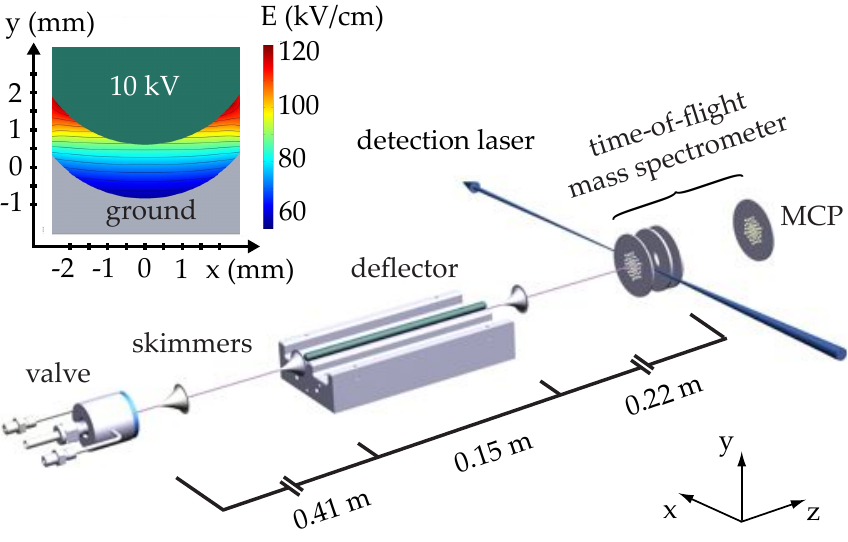}%
   \caption{Cold polar molecules in a supersonic jet are dispersed according to their effective
      dipole moments upon passing a strong inhomogeneous electric field. In the detection region
      quantum-state- and conformer-selective experiments can be conducted by changing the height of
      the detection laser. See text for details.}
   \label{fig:experiments:3AP_setup}%
\end{figure}

In all experiments discussed here, strong inhomogeneous electric fields are used to manipulate the
motion of large neutral molecules. In the conceptually most simple setup, depicted
in~\autoref{fig:experiments:3AP_setup}, a static electric field is used to disperse beams of polar
molecules in a deflector.
The experimental setup is described in detail elsewhere~\cite{Holmegaard:PRL102:023001,
   Filsinger:JCP131:064309}. In brief, a molecular beam is formed by expanding a mixture of helium
(50-100~bar) and the target molecules (at a partial pressure of a few mbar) through a pulsed
Even-Lavie valve~\cite{Hillenkamp:JCP118:8699} into vacuum. During the supersonic expansion the
molecules are efficiently cooled via collisions with the carrier gas to a rotational temperature of
$\sim$~1~K. The cold molecular beam is collimated using two skimmers before entering a 15-cm long
electrostatic deflector. A cut through the electric field of the deflector is shown in the inset
of~\autoref{fig:experiments:3AP_setup}. This so-called two-wire field~\cite{Ramsey:MolBeam:1956} has
a large gradient along the vertical $y$-axis and is homogeneous along the horizontal $x$-axis. Thus
polar molecules are predominantly deflected vertically with molecules in high-field-seeking quantum
states being deflected upwards. The deflected molecules then pass a third skimmer before they are
intersected in the interaction region by a focused ionizing laser pulse. The height of the detection
laser focus is scanned in order to measure the vertical molecular beam intensity profile. The
created ions, mass-selected by a linear time-of-flight mass spectrometer, are detected using a
microchannel plate detector.

\begin{figure}
   \centering%
   \includegraphics[width=\figwidth]{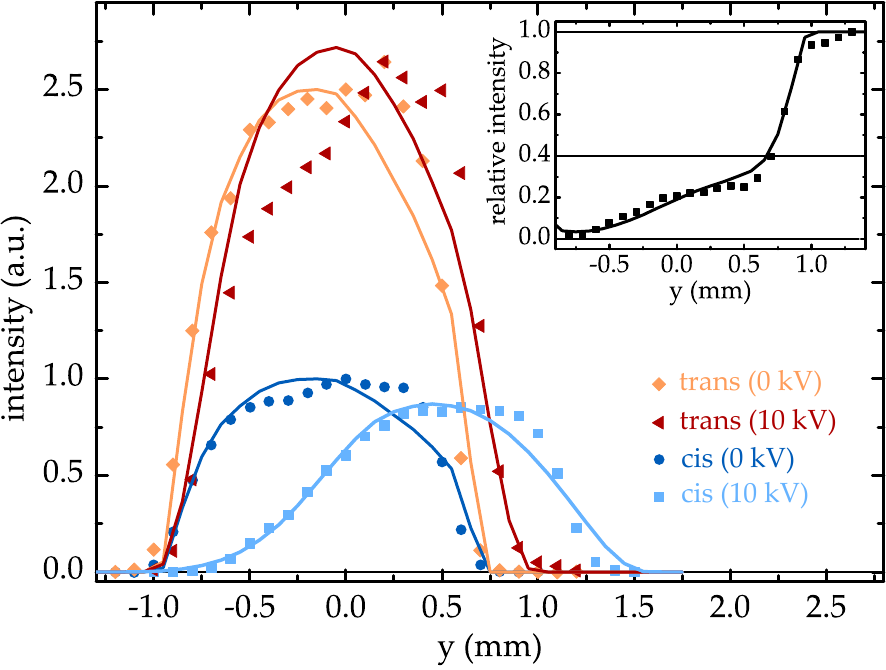}%
   \caption{Vertical molecular beam intensity profiles for cis-3AP and trans-3AP with and without
      high voltages applied to the deflector. Experimental data is shown as symbols, simulations are
      shown as solid lines. In the inset the fractional intensity of the cis conformer is shown as a
      function of the vertical position $y$ of the detection laser (reproduced from
      reference~\citealp{Filsinger:ACIE48:6900}).}
   \label{fig:experiments:3AP_deflection}%
\end{figure}

The idea behind this deflection setup is straightforward: upon passing through the electrostatic
deflector the species in the molecular beam are dispersed according to their effective dipole
moments $\mueff$. High-field-seeking molecules with a large and positive $\mueff$ end up very high
in the detection region. Unpolar species, such as the carrier gas atoms or molecules in very
high-lying rotational quantum states, will remain close to the molecular beam axis since they are
not deflected. If molecules in low-field seeking states (\ie, states with $\mueff < 0$) are present,
these molecules will reach the detection region below the molecular beam axis. We have demonstrated
that this experimental setup can indeed be applied for the conformer separation of large
molecules~\cite{Filsinger:ACIE48:6900}. \autoref{fig:experiments:3AP_deflection} shows a measurement
of the vertical intensity profiles for cis- and trans-3AP with and without high voltages applied to
the deflector.
The two conformers can be detected individually due to their distinct excitation wavelength in the
one-color resonance-enhanced multiphoton ionization scheme that is used for the
detection~\cite{Unterberg:CP304:237}. Without high voltages applied to the deflector, both
conformers exhibit the same spatial distribution. When 10~kV are applied to the deflector all
molecules are deflected upwards, as all quantum states of both conformers are high-field seeking at
the relevant electric field strenghts (see \autoref{fig:3AP:starkshift}). However, the deflection is
stronger for the more polar cis-conformer and above $y$=1~mm a pure sample consisting of only
cis-3AP is obtained. These isolated samples of cis-3AP molecules consist exclusively of molecules in
the lowest rotational quantum states, which have the largest effective dipole moments. The cis-3AP
molecules in high-$J$ states have smaller effective dipole moments, comparable to those of the
low-$J$ states of the trans conformer. Therefore, in the region around $y$=0.75~mm, cis-3AP
molecules in high-$J$ states and trans-3AP molecules in low-$J$ states spatially overlap. Only in
the lowest part of the molecular beam, below $y$=-0.75~mm where the population of cis-3AP is
completely depleted, a clean sample of the trans conformer is obtained. Note, however, that in this
region the trans-3AP molecules are predominantly in high-$J$ states. These molecules are still
overlaid with the He atoms from the carrier gas, which is not affected by the electric field. In
order to isolate the trans-3AP molecules in the lowest rotational states from both the cis conformer
and the carrier gas, the deflection experiment can be performed with Ne as the carrier gas, thereby
optimizing the deflection amplitudes by lowering the molecular beam velocity. Under these conditions
the beam is practically devoid of cis-3AP and pure samples of the lowest, most polar states of
trans-3AP can be obtained~\cite{Filsinger:thesis:2010}.

\begin{figure}
   \centering%
   \includegraphics[width=\figwidth]{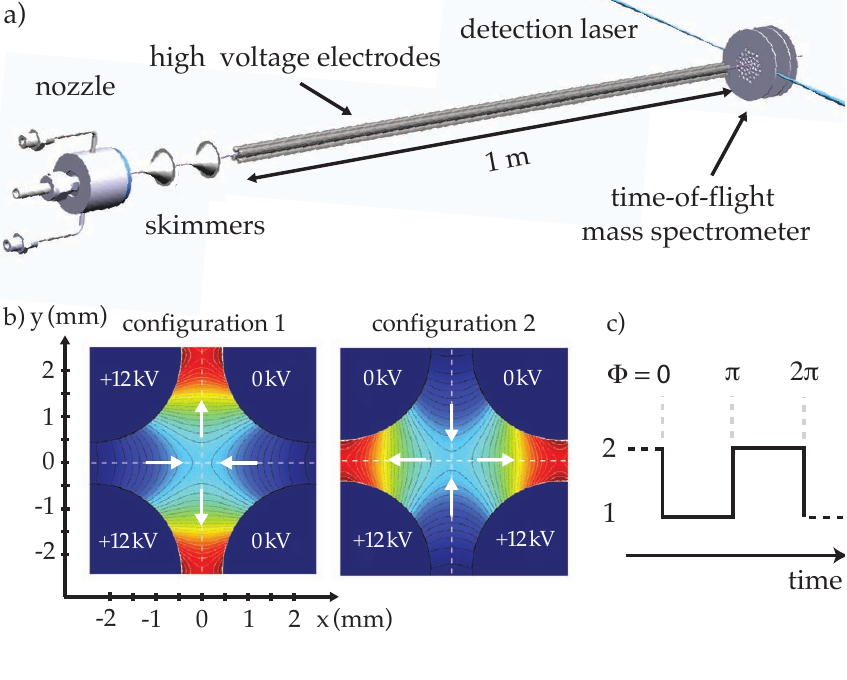}%
   \caption{(a) Scheme of the experimental setup used for alternating gradient focusing of
      high-field-seeking molecules. (c) Switching in a square wave pattern between the two electric
      field configurations shown in (b) results in a net focusing effect when the appropriate
      frequency is used.}%
   \label{fig:experiments:setup_selector}%
\end{figure}

One disadvantage of the electrostatic deflection technique is that it does not provide any focusing
but merely disperses the molecular beam. Many applications require tightly focused laser beams in
order to achieve the necessary light intensities. In these cases, a focused molecular beam would be
beneficial for optimal spatial overlap between the molecular sample and the detection laser. Whereas
small molecules in low-field-seeking quantum states can be focused using static multipole fields,
alternating gradient focusing is required to confine large molecules, which are high-field seeking
(see \autoref{sec:intro}). Electric fields suitable for AG focusing can be created
most easily by placing four cylindrical high voltage electrodes symmetrically around the molecular
beam axis and applying voltages as shown in~\autoref{fig:experiments:setup_selector}~b.
In the saddle-point like electric field of configuration 1 molecules are focused towards the
molecular beam axis along the $x$-direction and defocused along the $y$-direction. Switching to
configuration 2, which corresponds to configuration 1 rotated by 90\degree, interchanges focusing
and defocusing directions. The force acting on the molecules increases with increasing distance from
the molecular beam axis. Because the molecules are on average further away from the molecular beam
axis in the focusing lens compared to the defocusing lens, a net focusing effect results. The
optimal switching frequency depends on the dipole moment to mass ratio. If the field is switched too
slowly the molecules are pushed out of the electrodes along the defocusing direction before they are
refocused by switching the high voltages. If the field is switched too rapidly the molecules see the
time-averaged potential, which is defocusing for high-field-seeking molecules. Only for a small
range of frequencies AG focusing works. In other words, by choosing the appropriate switching
frequency only species within a small range of dipole moment to mass ratios are transmitted. The
operational principle of the device is analogous to that of the quadrupole mass filter for ions,
where molecules are discriminated based on their distinct charge to mass ratios. We have
demonstrated that AG focusing can also be exploited to select the individual conformers of
3AP~\cite{Filsinger:PRL100:133003}. At high ac frequencies, \ie, around 3~kHz, predominantly the
more polar cis conformer is transmitted, whereas the less polar trans conformer is selected at lower
ac frequencies ($\sim$1.5~kHz). Furthermore, similar to the deflection setup, quantum-state
selectivity is achieved since the lowest quantum states for a given conformer can be focused best.
In the initial experiments shown in~\cite{Filsinger:PRL100:133003}, the selectivity was inferior
compared to the deflection experiments. However, the selectivity of the focusing setup can be
considerably improved by lowering the molecular beam temperature thereby maximizing the population
of the lowest rotational quantum states. Moreover, we have demonstrated the possibility to
experimentally increase the resolution $\Delta\mu/\mu$ of the selector by changing the duty cycle of
the square wave shown in
\autoref{fig:experiments:setup_selector}~(c)~\cite{Filsinger:dutycycle:inprep}. This is,
effectively, the same effect as adding a dc offset in a quadrupole mass
filter~\cite{Richards:PIREEA32:321, Richards:IJMSIP12:317}.

In principle, a similar separation can also be achieved in the time domain. It has been shown that
large molecules can be decelerated using alternating gradient
decelerators~\cite{Wohlfart:PRA77:031404}. In principle, the AG decelerator could be used for the
conformer and quantum-state selection, since the deceleration process is quantum-state selective and
thus intrinsically conformer selective. However, AG deceleration is technically more challenging and
so far no species with multiple conformers has been decelerated.

For diffraction experiments at XFELs conformer-selected samples, prepared by one of the techniques
described above, are highly desired because the presence of multiple structural isomers will
prohibit analysis of the diffraction patterns. Moreover, the ideal targets for diffraction
experiments are molecular samples that are also aligned or oriented in the laboratory frame. Here
alignment refers to confinement of a molecule-fixed axis along a laboratory-fixed axis, and
orientation refers to the molecular dipole moments pointing in a particular direction. Alignment can
readily be obtained by the interaction of molecules with strong ac (laser)
fields~\cite{Friedrich:PRL74:4623, Stapelfeldt:RMP75:543}. Orientation is typically achieved through
hexapole state-selection for small molecules~\cite{Reuss:StateSelection}, brute-force
orientation~\cite{Friedrich:Nature353:412, Loesch:JCP93:4779}, or applying mixed ac and dc electric
fields~\cite{Friedrich:JPCA103:10280, Sakai:PRL90:083001, Buck:IRPC25:583} We have recently
demonstrated that the quantum-state selected polar samples produced by the manipulation methods
described above allow the creation of strongly aligned and oriented molecular
ensembles~\cite{Holmegaard:PRL102:023001, Filsinger:JCP131:064309}. Similar experiments on
hexapole-state-selected NO molecules in lfs states were performed using ultrashort laser pulses and
moderately strong ac fields~\cite{Ghafur:NatPhys5:289}

\begin{figure}
   \centering%
   \includegraphics[width=\figwidth]{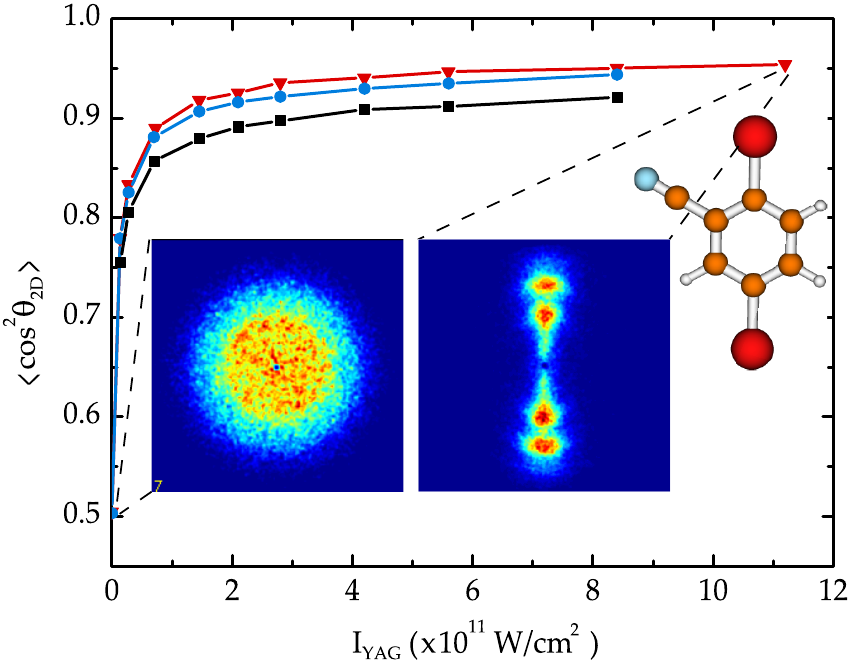}%
   \caption{Alignment of 2,5-diiodobenzonitrile as a function of the YAG laser intensity. Black
      squares are obtained for the undeflected molecular beam and blue circles (red triangles)
      correspond to a deflected sample at a height of 80\% (50\%) of the peak intensity in the
      deflection profile of the molecule obtained with 10 kV applied to the deflector. Inset: I$^+$
      images obtained with probe laser only (left) and with both alignment and probe laser (right).}
   \label{fig:25DIBN_alignment}%
\end{figure}

To illustrate the potential of this method, we studied adiabatic alignment of 2,5-diiodobenzonitrile
(DIBN) molecules, which are an interesting candidate for proof-of-principle X-ray diffraction
experiments. The experimental setup for these experiments is very similar to the one shown in
\autoref{fig:experiments:3AP_setup}, with the only difference that now two laser pulses intersect
the molecular beam in the interaction region. A 10~ns long YAG laser pulse is used to align the
molecules and a 30~fs long Ti:Sa laser pulse is employed to Coulomb explode the molecules at the
peak of the YAG laser pulse. Furthermore, a velocity map imaging spectrometer is used for detection.
The basic experimental observables, shown in \autoref{fig:25DIBN_alignment}, are 2D images of I$^+$
ions recorded when the molecules are irradiated with both the linearly polarized YAG pulse and the
fs probe pulse.
The angular distribution of the I$^+$ ions provides direct information about the spatial orientation
of the C-I bond axis of the DIBN molecules. When no YAG laser pulse is employed the ion distribution
is circular symmetric (see left inset of \autoref{fig:25DIBN_alignment}) and
$\langle\cos^2\theta_{2D}\rangle=0.50$, where $\theta_{2D}$ is the angle between the projection of
the I$^+$ recoil velocity on the detector plane and the YAG polarization. The image changes
dramatically when the YAG laser is employed. Now all I$^+$ are ejected in a narrow cone along the
laser polarization axis and $\langle\cos^2\theta_{2D}\rangle=0.95$ is observed for the highest laser
intensity and the most deflected molecules. From the measurements at different positions within the
vertical molecular beam intensity profile it is indeed clear that the degree of alignment
systematically increases with the deflection amplitude in consistency with previous
studies~\cite{Holmegaard:PRL102:023001, Filsinger:JCP131:064309}.

\section{Diffractive imaging of controlled molecular ensembles}
\label{sec:diffraction}

The conformer-selected and oriented molecular ensembles provided by state-of-the-art molecular beams
and the manipulation methods described above are ideal targets for diffractive imaging experiments
using novel femtosecond XFELs~\cite{Emma:NatPhoton, Altarelli:XFEL-TDR:2006} or ultrashort electron
packets~\cite{Siwick:JAP92:18857, Williamson:Nature386:159, Reckenthaeler:PRL102:213001}. If one
would take photographs of such ensembles, all molecules would look identical and they would also all
be in the identical pose. Such samples allow to directly measure molecular properties -- which
manifest themselves in the molecule-fixed frame -- in the laboratory space, \ie, the frame of
measurement: the link between the two frames of reference is given by the control over the motion of
the molecules demonstrated here. Possible applications of this involve the measurement of
photoelectron angular distribution functions of large molecules~\cite{Holmegaard:NatPhys6:428}, and
the search for associated interferences in ``diffraction from within''
measurements~\cite{Landers:PRL87:013002} (sometimes also called ``photoelectron
holography''~\cite{Szoeke:AIPCP147:361, Barton:PRL61:1356, Krasniqi:PRA81:033411}), which would
yield detailed information on the molecular structure. Other fields which could benefit from the
controlled samples include short-pulse dynamics, including impulsive alignment and the observation
of the full quantum structure of rotational dynamics, high-harmonic-generation and attosecond
experiments, stereochemically controlled photodissociation, half-collisons of weakly bound
complexes, or generally reaction dynamics.

Here, we discuss the prospects of these samples for experiments in which we more directly take
actual ``photographs'' of the molecules, \ie, we describe the possibility to apply these samples in
diffractive imaging experiments. We propose to acquire time-resolved images of aligned and oriented
individual structural isomers of molecules by coherent diffractive imaging (CDI) with XFEL pulses.
The exact investigations detailed below will likely not provide new insight into the structure and
dynamics of the relatively simple example molecules. However, they will benchmark the proposed
experimental techniques and evaluate the possibilities offered by the new instrumentation that is
becoming operational right now. This includes, for example, the structural biology experiments
described in \autoref{sec:intro:diffraction}.

\begin{figure}
   \centering%
   \includegraphics[width=\figwidth]{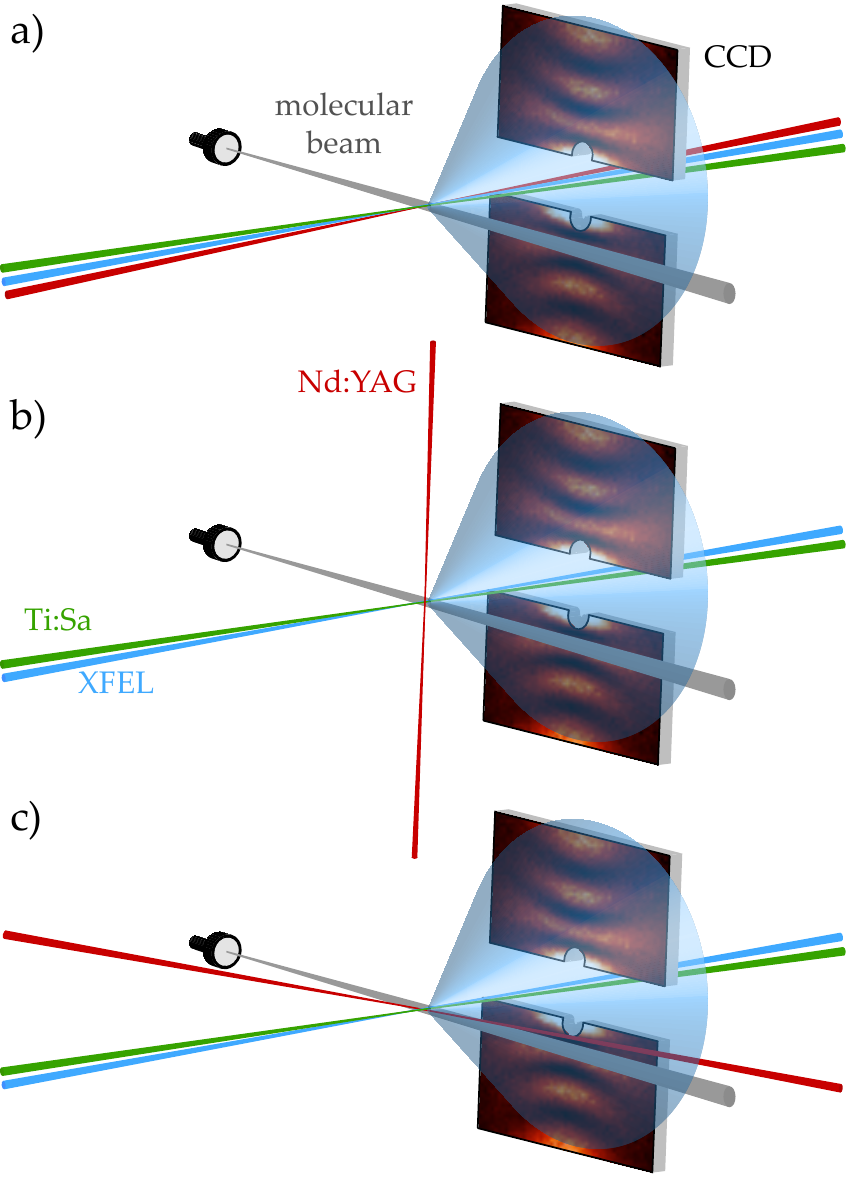}
   \caption{Sketch of three possible experimental setups for ultrafast diffractive imaging of
      aligned and oriented molecular ensembles. (a) The alignment laser (red), the pump laser
      (green), and the X-ray (probe) laser (blue) are all co-propagating, crossing the molecular
      beam at right angle, and continuing through an opening between the two panels of the CCD
      detectors. (b) The pump laser and the X-ray laser are co-propagating, crossing the molecular
      beam at right angle, and continuing through an opening in the CCD detector. However, in this
      setup the alignment laser crosses the molecular beam and the pump and probe lasers in the
      interaction region also at right angles. (c) The pump laser and the X-ray laser are
      co-propagating, crossing the molecular beam at nearly right angle (75\degree), and continuing
      through an opening in the CCD detector. However, in this setup the alignment laser is oriented
      in the plane of the other beams and crosses the molecular beam at a small angle (30\degree)
      and the pump and probe lasers in the interaction region at nearly right angles (-75\degree)}%
   \label{fig:diff:setup}%
\end{figure}

A sketch of the experimental setup is shown in \autoref{fig:diff:setup}.
In all experimental approaches one would use an electric deflector to spatially separate the (polar)
molecules from the atomic seed gas in order to mitigate the background that would result from
scattering of X-rays. It should be realized that the background scattering of helium from a seeded
molecular beam is comparable to the scattering of the seed molecules: while the scattering cross
section for helium is much smaller than for the molecules, there is a strong excess in particle
numbers on the order of $10^4$ helium atoms per molecule.

Generally, one would record the wide-angle forward-scattered diffraction pattern, from synchronized
pulsed packets of molecules in a molecular beam, on a CCD detector. Additionally, the interaction
region would be intersected by an alignment laser pulse and by an ultrashort laser pulse which are
also both synchronized to the XFEL pulses. Strong 1D and 3D alignment or orientation of the
molecular samples in space will be induced by the alignment laser pulses and dc electric
fields~\cite{Stapelfeldt:RMP75:543, Friedrich:JPCA103:10280, Holmegaard:PRL102:023001,
   Nevo:PCCP11:9912}. Comparing the different setups sketched in \autoref{fig:diff:setup} scheme~(a)
is the conceptionally simplest setup, but requires the merging of three laser beams, \ie, a pulsed
nanosecond near-infrared (NIR) laser for adiabatic alignment, a NIR, visible (VIS), or ultraviolet
(UV) ultrashort laser pulse inducing molecular dynamics, and the X-ray laser. Thiscl can be achieved
using a holey mirror, where the X-rays are transmitted through the hole and the optical wavelengths
are reflected by appropriate dielectric coatings on the mirror substrate. However, in this setup a
change of the polarization axis of the alignment laser, which rotates the molecular sample in space,
is identical to a rotation of the camera about the laser axis. Therefore, this setup does not allow
tomographic reconstruction experiments. This possibility can be obtained by changing the alignment
laser axis so that it does not coincide with the camera normal, as depicted in
\autoref{fig:diff:setup}~b and c. In both setups one would create a line-focus of the alignment
laser (micrometers in the focused dimension and millimeters in the unfocused dimension) in order to
align the full molecular ensemble probed by the X-ray laser, \ie, the intersection column between
the X-ray beam (with \micrometer{} diameter) and the molecular beam (with mm diameter).
\autoref{fig:diff:setup}~b would be the optimal setup regarding tomography, but would strongly
interfere with secondary detectors and diagnostics, \ie, ion and electron spectrometers. These
would, however, be necessary to determine and optimize spatial and temporal overlap of all pulses
and to determine and measure the degree of alignment of the molecular ensemble. Moreover, it would
be crucial to correlate the obtained diffraction images to molecular processes, for example, to the
radiation damage characterized by the observed charge states.

We point out that we are discussing adiabatic alignment, where the X-ray scattering occurs while the
molecule is in the strong ac field of the alignment laser. Laser-field-free aligned samples can be
obtained using femtosecond laser methods but typically the degree of alignment is considerably lower
compared to that obtained with adiabatic alignment~\cite{Felker:JPC90:724, RoscaPruna:PRL87:153902,
   Lee:PRL97:173001}. Also, field-free alignment occurs only in a narrow time window (few hundred
femtoseconds) which may be too short to follow a reaction through its entire duration. Moreover,
while the laser field could influence electronically excited states, the ground state structures of
molecules, discussed here, are not significantly influenced by the off-resonant laser field.
Therefore, there is no influence on the X-ray diffraction signal in the discussed limit where a
molecule interacts, on average, with less than one X-ray photon. This can be different for
considerably more intense X-ray pulses or for processes based on the absorption of UV/VIS/NIR or
X-ray-photons, where the created excited states could be influenced by the alignment laser field.
This includes, for example, studies of dynamics of electronically excited states or X-ray-ionization
photoelectron measurements. The influence of these effects on the proposed dynamics studies need to
be investigated, making the proposed benchmark experiments on well-known molecules even more
important.

The diffraction pattern of a single molecule is the intersection of the molecular transform, \ie the
continuous 3D Fourier transform of the molecules' electron density, with the Ewald sphere. The
diffraction pattern of an ensemble of gas-phase molecules is the incoherent sum of all the
individual patterns, even though with an XFEL the ensemble might be illuminated with a spatially
coherent beam. Thus the signal depends linearly on the number of molecules. The coherence of the
X-ray pulses will lead to speckles that are of size inversely proportional to the largest
intermolecular separations, but these will be much smaller than a single detector pixel and will be
averaged out. Additionally, the molecules are randomly positioned -- so diffracted amplitudes tend
not to sum in phase -- and the distribution changes shot to shot. For perfectly oriented ensembles
of molecules, the accumulated diffraction pattern would, therefore, be the incoherent sum of the
identical single-molecule patterns. Using the phase-retrieval algorithms of coherent diffractive
imaging one can reconstruct 2D images of those patterns, in a specific view. By varying the
alignment laser polarization (in setups \autoref{fig:diff:setup}~b and c), and hence the angle of
the alignment and orientation axis of the molecules with direction of the x-ray probe pulses, one
builds up the 3D molecular transform, which can be phased to give the molecules' 3D image. Using a
wavelength of \picometer{155} (\keV{8}) one can achieve an imaging resolution on the order of the
distances between neighboring atoms in a molecule.

\begin{figure}
   \centering%
   \includegraphics[width=\figwidth]{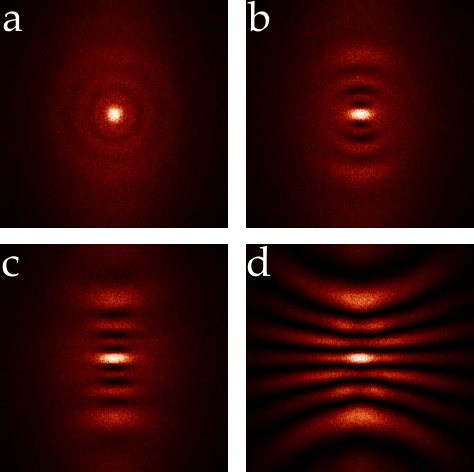}%
   \caption{Diffractive imaging patterns for 1D aligned ensembles of 2,5-diiodobenzonitrile at a
      X-ray energies of \keV{8} (\picometer{155}). (a) diffraction pattern of unaligned (isotropic)
      samples of DIBN b-d) diffraction patterns of 1D aligned and oriented samples of DIBN where
      degrees of alignment are \classTheta{20} (\cosTwoTheta{0.88}), \classTheta{10}
      (\cosTwoTheta{0.97}), and \classTheta{0} (``perfect'' alignment), for images b, c, and d,
      respectively, and the orientation is always according to an up:down ratio of 10:1. the
      scattering angle at the mid-point of the CCD detector edge is $2\theta=60\degree$.}
  \label{fig:diff:sim:DIBN:static:images}%
\end{figure}

In initial experiments one would investigate the wide-angle diffraction imaging of simple organic
molecules containing two iodine atoms, for example, the above mentioned 2,5-diiodobenzonitrile. Its
diffraction pattern will mainly consist of ``double slit'' like structures due to the two
electron-rich iodine centers. For DIBN the I--I separation of \unit[700]{pm} is on the order of the
currently available shortest wavelength of \unit[620]{pm} at LCLS, but ultrashort X-ray pulses at
shorter wavelength down to 100~pm will be available soon. In any case, clear fringes will be visible
in the diffraction images. In \autoref{fig:diff:sim:DIBN:static:images} simulated diffraction images
using \picometer{155} X-ray pulses of static samples of DIBN are shown for various degrees of 1D
alignment. The simulations include photon counting statistics and no instrumental noise.
In these simulations we have assumed a molecular density of $10^{10}$~cm$^{-3}$ and $10^{13}$ X-ray
photons/pulse focused to \micrometer{10} (95\,\% intensity diameter), resulting in $\sim\!160$
molecules in the 5~mm long interaction column of the X-ray beam and the molecular beam. Averaging
over $10^5$ pulses, corresponding to an acquisition time of 14~min at a repetition rate of 120~Hz as
available at LCLS, results in the given simulations.

\begin{figure}
   \centering
   \includegraphics[width=\figwidth]{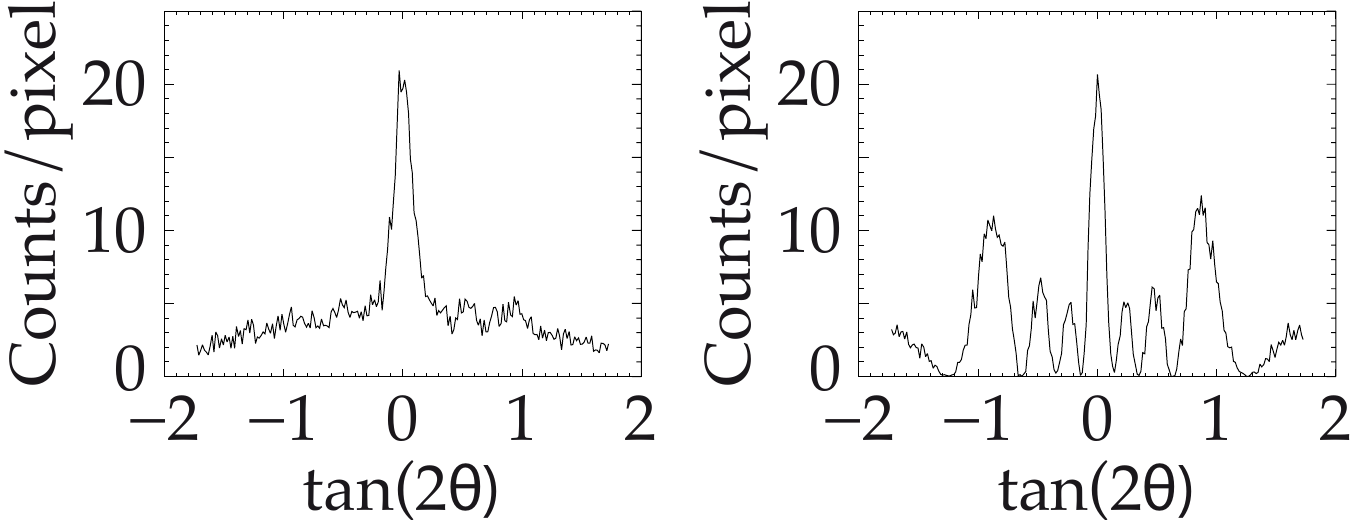}
   \caption{Vertical lineouts of the simulated diffraction pattern of (a) unaligned and (b)
      1D-aligned DIBN obtained from the images in \autoref{fig:diff:sim:DIBN:static:images} (a, d),
      averaged over a five pixel wide column.}
   \label{fig:diff:sim:DIBN:static:lineouts}
\end{figure}

In \autoref{fig:diff:sim:DIBN:static:images} simulated diffraction patterns for
2,5-diiodobenzonitrile are shown: In \autoref{fig:diff:sim:DIBN:static:images}~(d) for a
hypothetical (infeasible) perfectly 1D aligned and oriented ensemble (\cosTwoTheta{1}) and different
aligned samples with classical turning points for the alignment cone of (c) \classTheta{10}
($\cosTwoTheta{0.97}$), (b) \classTheta{20} (\cosTwoTheta{0.88}), and (d) an isotropic ensemble are
shown.
From these images it is obvious
that the contrast in such diffraction experiments will tremendously benefit from strongly aligned
samples, \ie, it would be extremely helpful to achieve molecular alignment of
$\left<\cos^2\theta\right>>0.9$ in order to obtain maximum fringe contrast for the central vertical
lineouts of these images as shown in \autoref{fig:diff:sim:DIBN:static:lineouts}.
From the spacing (in reciprocal space) of two minima or maxima in the lineouts one can directly
derive the dimensions of the molecule, \ie, the I--I distance in real space, by inversion of Bragg's
condition. For such a simple molecule, the determined heavy-atom distance can be compared to
quantum-chemical calculations and spectroscopic data, providing a detailed benchmark on the
feasibility of precise structure determination of larger molecules using XFEL radiation.

\begin{figure}
   \centering%
   \includegraphics[width=0.66\figwidth]{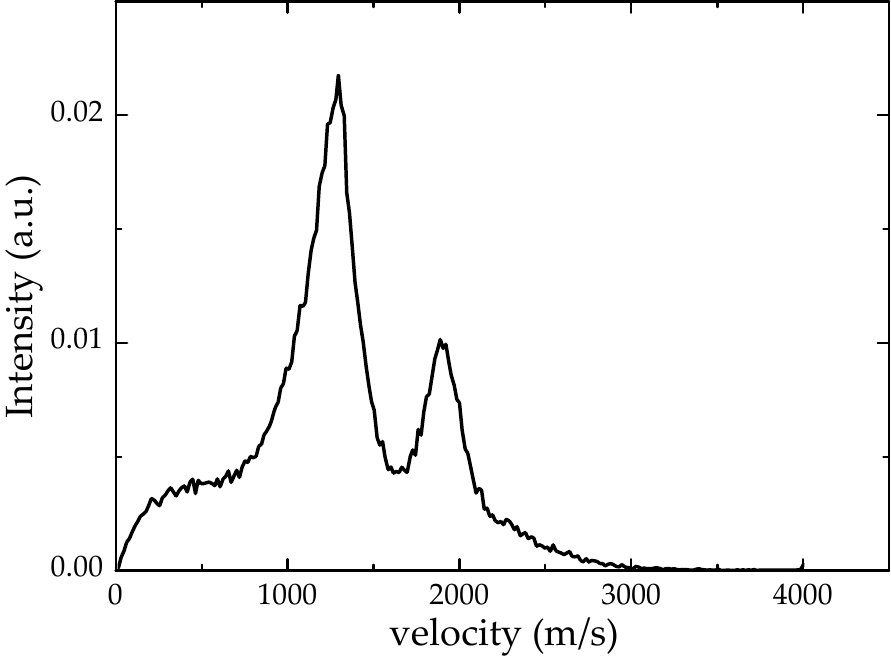}%
   \caption{Iodine ion velocity distributions obtained from Coulomb explosion imaging of
      diiodobenzene~\cite{Kumarappan:JCP125:194309}.}
    \label{fig:dissociation:velocity}%
\end{figure}

\begin{figure}[t]
   \centering%
   \includegraphics[width=\figwidth]{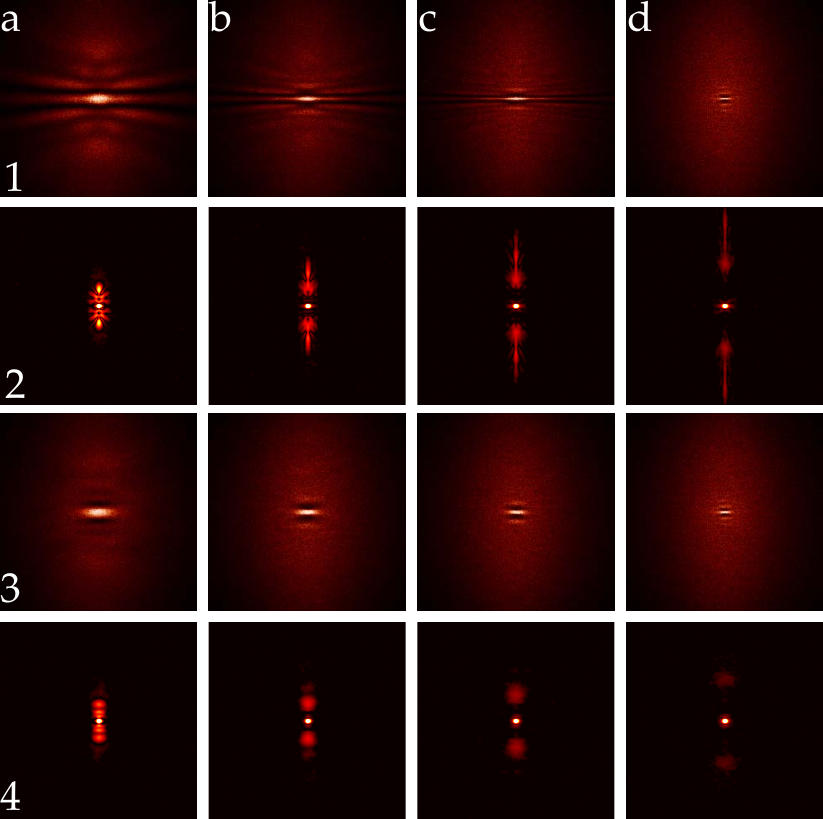}%
   \caption{Diffractive imaging patterns and their Fourier transforms for dissociating 1D aligned
      ensembles of 2,5-diiodobenzonitrile at an X-ray energy of \keV{8} (\picometer{155}) for
      pump-probe delays of (a) 0~fs, (b) 250~fs, (c) 500~fs, and (d) 1~ps. Row~1 shows the
      diffraction patterns of perfectly aligned samples, and row~2 the corresponding Fourier
      transforms. Row~3 and 4 show the same images for alignment to a cone of $\pm10\degree$.}
    \label{fig:diff:sim:DIBN:dissociation:images}%
\end{figure}

In subsequent experiments one would investigate ultrafast dynamics, such as vibrational,
torsional~\cite{Madsen:PRL102:073007}, or dissociation dynamics of the molecules studied. Here, we
focus on dissociation dynamics, resulting in the largest structural changes. We assume that we
multiply-ionize DIBN using a few-fs off-resonant NIR pulse, which results in Coulomb explosion and
axial recoil of the two I$^+$ fragments with the velocity distribution shown in
\autoref{fig:dissociation:velocity}.
Using this distribution we can simulate the diffractive imaging patterns for various time-delays
between the dissociation pump pulse and the X-ray probe pulse using the same experimental parameters
as above. The resulting images for originally 1D aligned molecular ensembles of DIBN assuming
\cosTwoTheta{1} and \cosTwoTheta{0.97} (\classTheta{10}) are given in
\autoref{fig:diff:sim:DIBN:dissociation:images} rows 1 and 3, respectively.
Again, the spacings of the intensity minima and maxima in the diffraction patterns directly
translate into the I--I distance. The structures of images a1 and a3 are somewhat washed out due to
averaging over the timing jitter of the pump laser.
Moreover, rows 2 and 4 of \autoref{fig:diff:sim:DIBN:dissociation:images} show the corresponding
Fourier transforms of these diffraction patterns. For clarity the square roots of the amplitudes are
shown and the colorscale extends from the minimum value to one-half of the the maximum value. The
separations of the heavy iodine atoms are directly visible and the used velocity distribution from
\autoref{fig:dissociation:velocity} are clearly recognized. The images of the perfectly aligned
samples show rich structure which is washed out for the realistic case obtained for
\cosTwoTheta{0.97} and, especially, for longer pump-probe delays. Nevertheless, even for the longest
delay of 1~ps and the realistic degree of alignment one can clearly define the separation of the
major velocity component of the two main scattering centers. Determining these distances as a
function of pump-probe delay one can, therefore, directly follow the molecular motion in \emph{real
   time}.

In general, such experiments would fully utilize the properties of XFEL sources, \ie, the high peak
brightness and short wavelengths~\cite{Emma:NatPhoton, Altarelli:XFEL-TDR:2006}. The X-ray pulse
needs to be spatially coherent over the size of an individual molecule, but not over the size of the
sample, as the intensities of individual diffraction images are summed incoherently. The available
ultrashort pulses allow to obtain scattering data without blurring due to residual rotational or
induced vibrational/dissociative molecular motions, and the high pulse fluences allow to obtain
scattering signals from an ensemble of gas-phase molecules above experimental noise levels within
relatively short times. The short wavelengths of these XFELs are required to resolve the atoms in
molecular compounds. There have been many arguments and simulations on the concept of diffraction
before destruction~\cite{Neutze:Nature406:752}. It is generally understood that the intense pulses
from the XFELs will lead to very strong ionization of the samples and, successively, Coulomb
explosion of the molecules. However, if the X-ray pulses are short enough, all diffraction events
will be over by the time the molecule considerably changes its scattering factor. The observable
destruction depends on the X-ray fluence on the individual molecules and the length of the pulse.
For a given photon number per pulse the destruction will increase with decreasing focus
size~\footnote{In principle, similar investigations can be performed by attenuating the X-ray beam,
   which would, however, result not only in less destruction but also in similarly decreased signal
   intensities.}. Since the molecular beams are much larger than the interaction volume and have an
essentially uniform density on the scale of the X-ray focus, one can change the fluence without
affecting the overall scattering intensity, as for any linear process: While the fluence, and
therefore the scattering intensity per molecule, decreases quadratically with the focus diameter, at
the same time the number of molecules in the interaction volume increases quadratically. This
yields, nominally, the same diffraction intensity as long as the destruction is negligible. However,
as soon as the fluence is so high that is causes destruction on the time-scale of the X-ray pulse
duration, a decrease in the focus size will lead to a smearing out of the diffractive imaging
patterns due to summing over non-equivalent molecular systems.


\section{Summary}
\label{sec:summary}

Using static inhomogeneous electric fields, complex polar molecules can be deflected and spatially
dispersed according to their effective dipole moment, \ie, according to their quantum state. Using
switched electric fields one can also actively focus or even decelerate packets of molecules in a
small set of quantum states. Both approaches, deflection and focusing, can be used to prepare
packets of individual structural isomers of such complex molecules. Moreover, because the methods
intrinsically create very polar samples, these molecular ensembles can be aligned and oriented
extremely well. Overall, these techniques allow to prepare packets of individual structural isomers
that are all fixed in space due to large degrees of alignment and orientation. In addition, the
electric deflection allows for complete separation of the molecular ensemble from the atomic seed
gas, resulting in pure molecular samples.

We described how these samples can be exploited for diffractive imaging experiments, exemplified for
ultrafast X-ray diffraction using XFELs. These experiments provide detailed benchmarks on the
feasibility of coherent hard X-ray diffractive imaging of complex gas-phase molecules and the
diffract-before-destruct concept. The correctness of the extracted structural data can be compared
to independently determined spectroscopic results and X-ray diffraction data from crystals. By
changing the pulse length and the focus size of the X-ray pulses one can perform detailed studies of
the radiation damage and its influence on the observed diffractive imaging patterns. Moreover, it
will be possible to use these relatively simple diffraction patterns to test and calibrate
correlation and inversion algorithms necessary for the extraction of structures of larger molecules
from their diffractive imaging patterns. These experiments thus explore new paradigms in structure
determination that are enabled by XFELs. Such studies will provide a path to the imaging of peptides
and other complex molecules without the need for crystallization.

Moreover, we note that a number of complementary experiments would similarly benefit from the
controlled samples described above. This includes experiments on molecular frame photoelectron
angular distributions~\cite{Landers:PRL87:013002, NugentGlandorf:PRL87:193002,
   Holmegaard:NatPhys6:428}, including photoelectron holography~\cite{Szoeke:AIPCP147:361,
   Barton:PRL61:1356, Krasniqi:PRA81:033411}, high-harmonic generation and molecular orbital
tomography~\cite{Itatani:Nature432:867} or electron diffraction~\cite{Williamson:Nature386:159,
   Siwick:Science302:1382}.

\begin{acknowledgments}
   We thank our colleagues at the Fritz Haber Institute and the University of Aarhus for many
   helpful discussions and Lotte Holmegaard, Jonas L.\ Hansen, Jens H.\ Nielsen, and Sofie Louise
   Kragh for help with obtaining the data in \autoref{fig:25DIBN_alignment}. Expert technical
   support at the Fritz Haber Institute and within the Department of Molecular Physics is gratefully
   acknowledged.
\end{acknowledgments}

\bibliographystyle{rsc}
\bibliography{string,mp,missing}

\end{document}